\documentclass[twocolumn,showpacs,preprintnumbers,amsmath,amssymb]{revtex4}

\usepackage{graphicx}
\usepackage{dcolumn}
\usepackage{bm}
\usepackage[usenames]{color}
\definecolor{Blue}{rgb}{0,0.08,0.65}
\definecolor{Red}{rgb}{0.65,0.08,0.05}
\definecolor{Green}{rgb}{0.15,0.45,0.25}

\begin{document}


\title{ Non Gaussian extrema counts for CMB maps 
}

\author{Dmitri Pogosyan${}^{1}$ }
\author{Christophe Pichon${}^{2}$ }%
\author{Christophe Gay${}^{2}$ }
\affiliation{%
${}^{1}$ Department of Physics, University of Alberta, 11322-89 Avenue, Edmonton, Alberta, T6G 2G7, Canada\\
${}^{2}$ Institut d'astrophysique de Paris,  98, bis boulevard Arago, 75 014, Paris, France
}
%
%
\begin{abstract}
In the context of the geometrical analysis of weakly non Gaussian CMB maps,
the 2D differential extrema counts as functions of the excursion set threshold
 is derived from the full moments expansion of the joint probability distribution
of an isotropic random field, its gradient and invariants of the Hessian. Analytic expressions for these counts
are given to second order in the non Gaussian correction, while a Monte Carlo method to compute them to arbitrary order is presented.
Matching count statistics to these estimators is illustrated on fiducial non Gaussian  ``Planck" data.
\end{abstract}


\maketitle

Random fields are ubiquitous phenomena in physics appearing in areas  ranging from
turbulence to the landscape of string theories.
In cosmology, the sky-maps of the polarized Cosmic Microwave Background (CMB) radiation 
-- a focal topic of current research --
is a prime example of  such  2D random fields.  
Modern view of the cosmos, developed primarily through statistical
analysis of these fields, points to a Universe that is 
statistically homogeneous and isotropic with a hierarchy of structures 
arising from small Gaussian fluctuations of quantum origin.
While the Gaussian limit provides the fundamental starting point in the study
of random fields \cite{Adler,Doroshkevich,BBKS},
non-Gaussian features of the CMB fields are of great interest.
Indeed, CMB inherits a high level of gaussianity from 
initial fluctuations, but small non-Gaussian deviations may
provide a unique window into the details of processes in the early Universe.
The search for the best methods to analyze 
non-Gaussian random fields  is ongoing.  

In paper \cite{PGP} the general invariant based formalism for computing
topological and geometrical 
characteristics of non Gaussian fields was presented. The general formulae for
the Euler characteristics 
to all orders has been derived, which encompasses the well known first correction \cite{Matsubara} 
and which was later confirmed to  the next order by \cite{matsu10}. 
We now focus on the statistics of 
the density of extremal points which follows directly from the formalism of \cite{PGP}.
The goal of this paper is to provide an explicit recipe on how to use this formalism in practice on idealised
2D CMB ``Planck"-like data.

\section{Extrema counts} 
Extrema counts, especially that of the maxima of the field, have long application to cosmology
\cite[e.g.][]{BBKS},
however  theoretical development  have been mostly restricted
to the Gaussian fields.  The statistics of extrema counts, as well as of the
Euler number, requires the knowledge of the one-point JPDF $P(x,x_i,x_{ij})$
of the field $x$, its
first, $x_i$, and second, $x_{ij}$, derivatives \footnote{\label{sigmas}We
consider the
field and its derivatives to be normalized by their corresponding variances
$\sigma_0^2=\left\langle x^2 \right\rangle, \sigma_1^2 = \left\langle (\nabla
x)^2
\right\rangle, \sigma_2^2 = \left\langle (\Delta x)^2 \right\rangle $.
This implies that in our dimensionless units $\langle x^2 \rangle=\langle
q^2 \rangle=\langle J_1^2 \rangle=\langle J_2 \rangle=1$.
}.
Extrema density is an intrinsically isotropic statistics given by
\citep{Adler,Longuet}
\begin{equation}
\frac{\partial n_{\rm ext}}{ \partial x} = \int {\rm d}^6 x_{ij}
P(x,x_i=0,x_{ij})
|x_{ij}| \,.
\label{eq:ext_int}
\end{equation}
Under
the condition of statistical isotropy of the field, the essential form
for the JPDF is therefore given in terms of the rotation invariants --- $x$
itself, the square of the
magnitude of the gradient $q^2\equiv x_1^2+x_2^2$ and the two invariants
$J_1\equiv\lambda_{1}+\lambda_{2}$,  $J_2\equiv(\lambda_{1}-\lambda_{2})^2$
of the Hessian matrix $x_{ij}$ (where $\lambda_i$ are the
eigenvalues of the Hessian). Introducing  $\zeta=(x+\gamma
J_1)/\sqrt{1-\gamma^2}$ (where the spectral parameter $\gamma=-
  \langle x J_1 \rangle$ 
characterizes the shape of the underlying power spectrum),
 leads to the following
JPDF for the Gaussian 2D field
\begin{equation}
G_{\rm 2D} = \frac{1}{2 \pi} 
\exp\left[-\frac{1}{2} \zeta^2 - q^2 - \frac{1}{2} J_1^2 - J_2 \right] \,.
\label{eq:2DG}
\end{equation}
The invariant form for the extrema counts
\begin{equation}
\frac{\partial n_{\rm ext}}{\partial x} =
\int \!\!\! \frac{{{\rm d} J_1} {{\rm d} J_2}}{8\pi^2 \sqrt{1\!-\!\gamma^2}}
\exp\left[-\frac{1}{2} \zeta^2 - \frac{1}{2} J_1^2 -J_2\right]
\left|J_1^2-J_2 \right| \nonumber
\end{equation}
then readily recovers the classical results  \citep{Adler,Longuet,BBKS}
when the limits of integration
that define the extrema type are implemented, namely
$J_1 \in [-\infty,0]$, $J_2 \in [0, J_1^2] $ for maxima, $J_1 \in
[0,\infty]$, $J_2 \in [0, J_1^2] $ for minima and $J_1 \in
[-\infty,\infty]$,$J_2 \in [J_1^2, \infty]$  for saddle points.

In \cite{PGP} we have observed that for non-Gaussian  JPDF the invariant
approach immediately suggests a Gram-Charlier expansion  in
terms of the orthogonal polynomials defined by the kernel $G_{\rm 2D}$.
Since $\zeta$, $q^2$, $J_1$ and $J_2$ are uncorrelated variables in the
Gaussian limit,
the resulting expansion is
\begin{widetext}
\begin{equation}
P_{\rm 2D}(\zeta, q^2, J_1, J_2) =  G_{\rm 2D} \left[ 1 + 
\sum_{n=3}^\infty \sum_{i,j,k,l=0}^{i+2 j+k+2 l=n} 
\frac{(-1)^{j+l}}{i!\;j!\; k!\; l!} 
\left\langle \zeta^i {q^2}^j {J_1}^k {J_2}^l \right\rangle_{\rm GC}
H_i\left(\zeta\right) L_j\left(q^2\right)
H_k\left(J_1\right) L_l\left(J_2\right)
\right]\,,
\label{eq:2DP_general}
\end{equation}
\end{widetext}
where terms are sorted in the order of the field power $n$
and $\sum_{i,j,k,l=0}^{i+2 j+k+2 l=n} $
stands for summation over all combinations
of non-negative $i,j,k,l$ such that $i+2j+k+2l$ adds 
to the order of the expansion term $n$. 

The Gram-Charlier coefficients, 
$\left\langle \zeta^i {q^2}^j {J_1}^k {J_2}^l \right\rangle_{\rm GC}\equiv
(-1)^{j+l} j! l! \left\langle H_i\left(\zeta\right) L_j\left(q^2\right)
H_k\left(J_1\right) L_l\left(J_2\right) \right\rangle_{\rm m}$
that appear in the expansion can be related to the more familiar cumulants
of the field and its derivatives (we use $\langle \,\,\,\,\rangle_{\rm m}$ for
statistical moments while reserving $\langle \,\,\,\,\rangle$ for statistical
cumulants), actually being identical to them for the first three orders
$n=3,4,5$. Lookup tables of the relationship between Gram-Charlier cumulants
and statistical  cumulants
 can be found at {\tt http://www.iap.fr/users/pichon/Gram/}. As an illustration,
one
sixth order non trivial cumulant would be 
 $\left\langle J_1^3 J_2 \zeta
   \right\rangle
   {}_{\text{CG}}=\left\langle J_1^3 J_2 \zeta
   \right\rangle+\left\langle
   J_1^3\right\rangle  \left\langle
   J_2 \zeta \right\rangle +3
   \left\langle J_1
   J_2\right\rangle  \left\langle
   J_1^2 \zeta \right\rangle
   $. 
It is prudent to stress that the Gram-Charlier series expansion is distinct from the
perturbative  expansions. For instance, while the linear Edgeworth or $f_{\rm NL}$
expansion  match solely to the first order $n=3$ Gram-Charlier
coefficients, quadratic terms require knowledge of the Gram-Charlier terms to
$n=6$, while the cubic ones to $n=9$. 

Integrals over $J_1$ and $J_2$ for extremal points
can be carried out analytically even for the
general expression (\ref{eq:2DP_general}).
Different types of critical points can be evaluated separately
by restraining the integration domain in the $J_1$-$J_2$ plane to ensure the
appropriate signs for the eigenvalues.

The effect of the non-Gaussian cubic correction on the total number of the
extrema of different types is given by
\begin{eqnarray}
n_{\rm max/min} &=&  \frac{1}{8 \sqrt{3} \pi {R_*}^2} \pm
\frac{18 \left\langle q^2 J_1 \right\rangle - 5\left\langle J_1^3 \right\rangle
+ 6 \left\langle J_1 J_2 \right\rangle}{54 \pi \sqrt{2\pi} {R_*}^2}
\,, \nonumber \\
n_{\rm sad} &=& \frac{1}{4 \sqrt{3} \pi{R_*}^2} 
\,,
\end{eqnarray}
where we have restored (see note [11]) 
the dimensional scaling with $R_* = \sigma_1/\sigma_2$ ,
the characteristic separation scale between  extrema.
The total number of saddles, as well as of all the extremal points, $n_{\rm max}
+ n_{\rm min} + n_{\rm sad}$, are
preserved in the first
order (the latter following for the former, as topological considerations imply
$n_{\rm max}-n_{\rm sad}+n_{\rm min}={\rm const}$), but the symmetry between the
minima and the maxima is  broken.

The differential number counts with respect to the excursion threshold $\nu$ are
given by
\begin{widetext}
\begin{eqnarray}
\frac{\partial n_{\rm max/min}}{\partial \nu} &=& 
 \frac{1}{\sqrt{2 \pi}{R_*}^2} \exp\left(-\frac{\nu^2}{2}\right) 
\left[1 \pm \mathrm{erf}\left(\frac{\gamma \nu}{\sqrt{2(1-\gamma^2)}}\right)
\right] K_1(\nu,\gamma) 
\pm \frac{1}{\sqrt{2 \pi (1-\gamma^2)}{R_*}^2}
\exp\left(-\frac{\nu^2}{2(1-\gamma^2)}\right)
K_3(\nu,\gamma) \nonumber
\label{eq:difnu1}
\\
&+& \frac{\sqrt{3}}{\sqrt{2 \pi (3-2\gamma^2)}{R_*}^2}
\exp\left(-\frac{3 \nu^2}{6-4 \gamma^2}\right) \left[1 \pm 
\mathrm{erf}\left(\frac{\gamma \nu}{\sqrt{2(1-\gamma^2)(3-2\gamma^2)}}\right)
\right] K_2(\nu,\gamma) ,
\\
\frac{\partial n_{\rm sad}}{\partial \nu} &=& 
\frac{2 \sqrt{3}}{\sqrt{2 \pi (3-2\gamma^2)}{R_*}^2}
\exp\left(-\frac{3 \nu^2}{6-4 \gamma^2}\right) K_2(\nu,\gamma), \label{eq:difnu2}
\end{eqnarray}
where $K_1, K_2, K_3$ are polynomials with coefficients expressed in terms of the
cumulants.  Here we give explicit expressions for the first non-Gaussian order,
while the next order can be found at the above mentioned URL.

The term $K_1(\nu,\gamma)$
has a special role determining the Euler number $\chi(\nu)$
via ${\partial  \chi}/{\partial \nu} = {\partial}/{\partial \nu}
\left(n_{\rm max} + n_{\rm min} - n_{\rm sad}\right) =
\sqrt{{2}/{\pi}} \exp(-{\nu^2}/{2}) K_1(\nu,\gamma)$.
As such, its full expansion has been given in \cite{PGP}, Eq.~(7), and confirmed
to the second order in \cite{matsu10}. To the leading non-Gaussian order
\begin{equation}
K_1 = \frac{\gamma^2}{8\pi}
\left[ H_2(\nu) + \left( \frac{2}{\gamma} \left\langle q^2 J_1
\right\rangle + 
\frac{1}{\gamma^2} \left\langle x {J_1}^2 \right\rangle - 
\frac{1}{\gamma^2} \left\langle x J_2 \right\rangle \right) H_1(\nu)
- \left( \left\langle x q^2 \right\rangle + 
\frac{1}{\gamma} \left\langle x^2 J_1 \right\rangle \right) H_3(\nu)
+ \frac{1}{6} \left\langle x^3 \right\rangle H_5(\nu) \right]\,.
\label{eq:2D_K3}
\end{equation}

Introducing scaled Hermite polynomials
${\cal H}_n^{\pm}(\nu,\sigma)\equiv \sigma^{\pm n} H_n\left(\nu/\sigma\right)$,
the polynomial $K_2(\nu,\gamma)$, the only one that determines the distribution
of saddle points, can be written as
\begin{eqnarray}
K_2 =\frac{1}{8\pi \sqrt{3}}
\left[\  
\vphantom{\frac{1}{6}} 1  \right.
- \left( \left\langle x q^2 \right\rangle 
+ \frac{1}{3} \left\langle x {J_1}^2 \right\rangle
- \frac{4}{3} \left\langle x J_2 \right\rangle
+ \frac{2}{3} \gamma \left\langle q^2 J_1 \right\rangle 
+ \frac{2}{9} \gamma \left\langle {J_1}^3 \right\rangle  
- \frac{2}{3} \gamma \left\langle J_1 J_2 \right\rangle \right)
{\cal H}_1^- \left(\nu,\sqrt{1-2/3\gamma^2} \right) \nonumber \\ 
\left. + \frac{1}{6} \left(
\left\langle x^3 \right\rangle
+ 2 \gamma \left\langle x^2 J_1 \right\rangle
+ \frac{4}{3} \gamma^2 \left\langle x {J_1}^2 \right\rangle
+ \frac{2}{3} \gamma^2 \left\langle x J_2 \right\rangle
+ \frac{8}{27} \gamma^3 \left\langle {J_1}^3 \right\rangle
+ \frac{4}{9} \gamma^3 \left\langle J_1 J_2 \right\rangle
\right)
{\cal H}_3^- \left(\nu,\sqrt{1-2/3\gamma^2} \right) \right].
\label{eq:2D_K2}
\end{eqnarray}

The remaining term, $K_3(\nu,\gamma)$ is the most complicated one. It is 
expressed as the expansion in ${\cal H}_n^+(\nu,\sqrt{1-\gamma^2})$:
\begin{eqnarray}
\lefteqn{K_3 =\frac{(1-\gamma^2)}{2 (2 \pi)^{3/2}(3-2\gamma^2)^3}
\left[\vphantom{\frac{1}{6}} \gamma (3-2\gamma^2)^3 {\cal H}_1^+
\left(\nu,\sqrt{1-\gamma^2}\right)
+ \left( 
\frac{1}{2}\gamma^3 \left(1+\gamma^2-26 \gamma^4 +28 \gamma^6 - 8 \gamma^8
\right)
\left\langle x ^3\right\rangle
\right.\right.}
\nonumber \\
&&\left.
-\gamma^4 \left(26-28 \gamma^2+8 \gamma^4\right)
\left\langle x^2 J_1 \right\rangle
+ \gamma \left(1-\gamma^2\right)\left(1+2 \gamma ^2\right) \left(3-2 \gamma
^2\right)^2
\left\langle x q^2 \right \rangle 
-\gamma \left(24-26 \gamma^2+8 \gamma^4 \right)
\left\langle x  J_1^2\right \rangle
\right. \nonumber \\
&&\left.\vphantom{\frac{1}{6}}
+\gamma\left(15-23 \gamma^2 + 8 \gamma^4 \right)
\left\langle x J_2\right \rangle
+ 4 (1-\gamma^2) \left(3 -2 \gamma ^2 \right)^2
\left\langle q^2 J_ 1\right \rangle
-\left(10 - 12 \gamma^2 + 4 \gamma^4 \right)
\left\langle J_1^3\right \rangle
+6 \left(1-\gamma^2\right) \left(2-\gamma^2\right)
\left\langle J_1 J_2\right \rangle
\right)
\nonumber \\
&& -\frac{1}{6} \left(\vphantom{\frac{1}{6}}
\gamma \left( 
27+36 \gamma^2-224 \gamma^4+192 \gamma^6-48 \gamma^8 \right)
\left\langle x^3\right \rangle 
+ \left( 108 - 324 \gamma^2+ 216 \gamma^4 - 48\gamma ^6 \right)
\left\langle x^2 J_1\right \rangle 
+ 6 \gamma (3 - 2 \gamma^2)^3 \left\langle x q^2 \right \rangle 
\right. \nonumber \\
&&\left. \left. \vphantom{\frac{1}{6}}
-36 \gamma \left\langle x J_1^2\right \rangle
-18 \gamma \left\langle x J_2\right \rangle
- 8 \gamma^2 \left\langle J_1^3\right \rangle 
- 12 \gamma^2 \left\langle J_1 J_2\right \rangle \right)
{\cal H}_2^+\left(\nu,\sqrt{1-\gamma^2}\right) \right]. \label{eq:2D_K1}
\end{eqnarray}
\end{widetext}
Eqs~(\ref{eq:difnu1})-(\ref{eq:difnu2}) (together with the next order expansion available online) are the main theoretical result of this paper.
 
\section{Implementation}

\begin{figure*}
\includegraphics[width=0.45\textwidth]{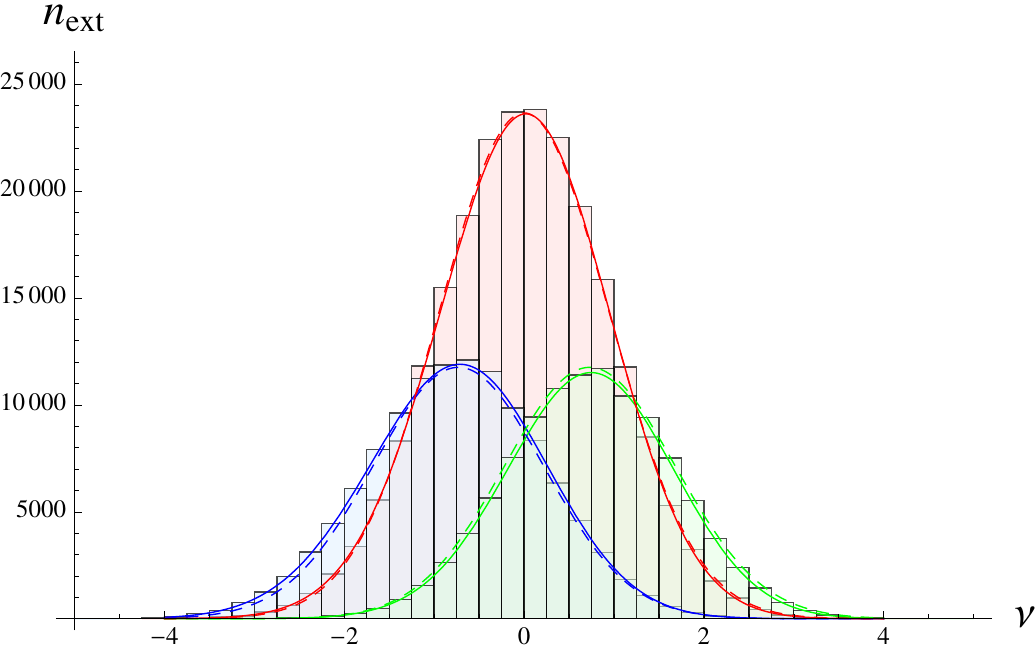}
\includegraphics[width=0.45\textwidth]{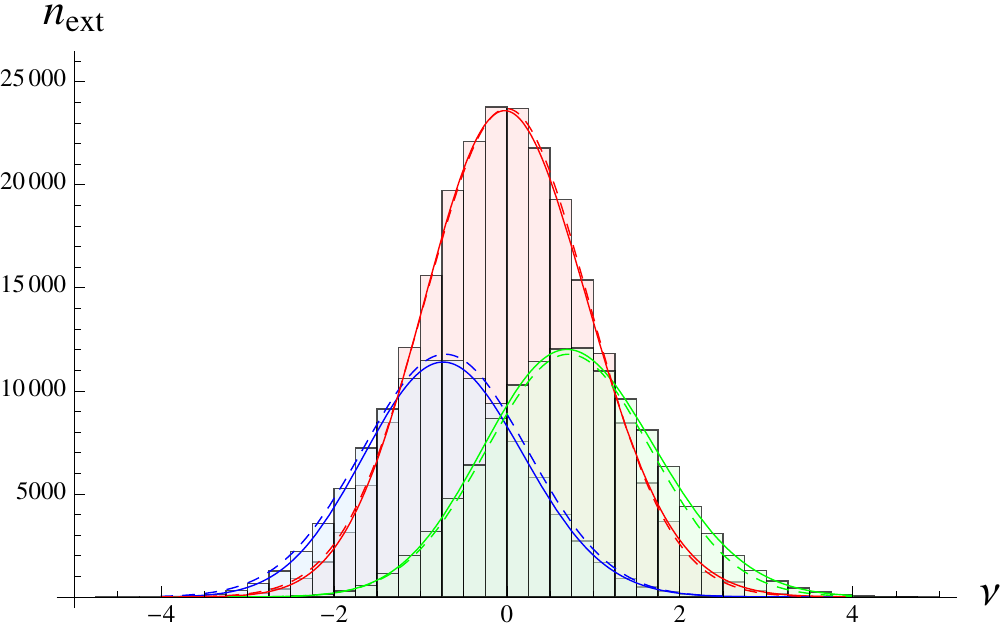}
\includegraphics[width=0.45\textwidth]{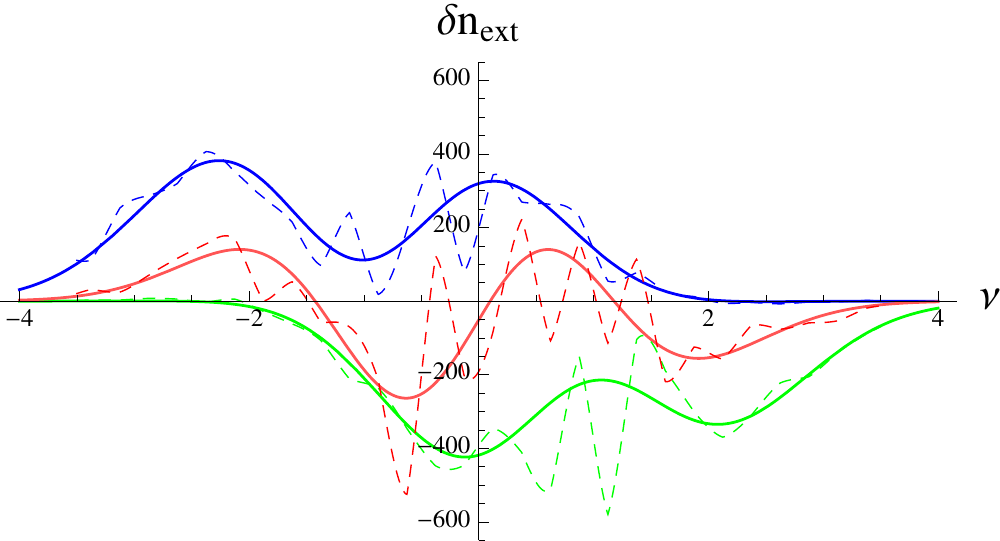}
\includegraphics[width=0.45\textwidth]{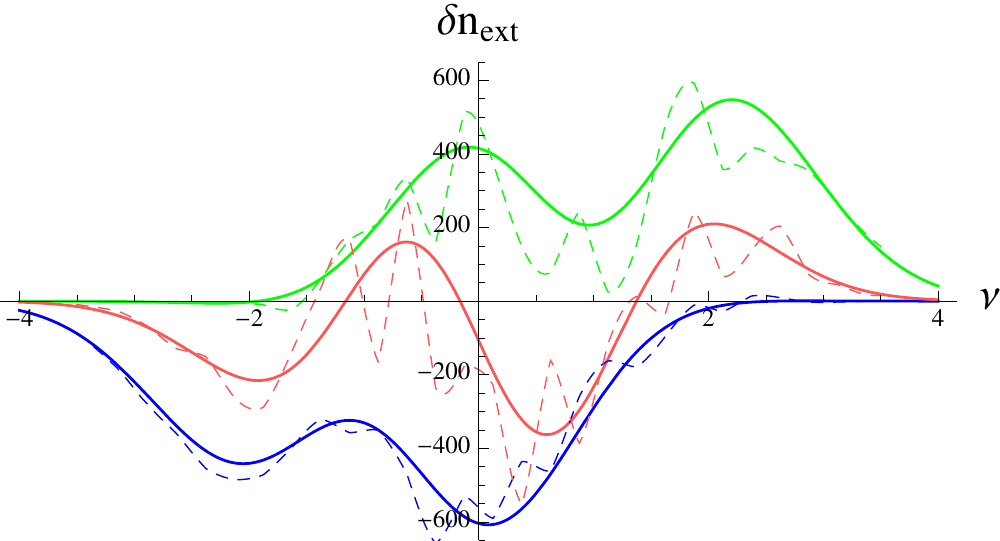}
\caption{{\sl Top panel}: the predicted ({\sl solid line}) number of
maxima, saddle, and minima in $\Delta\nu=0.25$ bins as a function of the
threshold, $\nu$, on top of the measured
count from a {\sl single} realization full-sky {\sc \small nside}=2048 {\sc \small
HEALPix} map ({\sl histogram}). The temperature field is smoothed with
the Gaussian filter of 10 arcmin FWHM, resulting in $R_* \approx 5.5$ arcmin
$\approx 3$ pixels. 
The dashed line corresponds to the Gaussian
prediction.
 The left panel corresponds to  the  Harmonic Oscillator model of non
Gaussianity with $\alpha_1=0.6$, $\alpha_2=0.6$,
while the right panel corresponds to the power law  non Gaussianity with $\beta=2$.
{\sl Bottom panel}: the departure from Gaussianity for these two models as
predicted ({\sl solid line}) and measured ({\sl dashed line}) for maxima ({\sl green}), minima
({\sl blue}) and saddle points ({\sl red}).
 Note that the corrections of Eqs~(\ref{eq:difnu1})-(\ref{eq:difnu2}) ({\sl solid
line}) give a very accurate match to the measured PDF.
As is seen,
different models of non-Gaussianity can be distinguished by their effects on
extrema.
} 
\label{fig:mainfig}
\end{figure*}

Evaluating these estimators requires computing the cumulants appearing
in Eqs.~(\ref{eq:2D_K3})-(\ref{eq:2D_K1}).
In non-Gaussian models where the field is represented by the functional of 
a Gaussian field this may be possible directly, while in general, 
as shown in \cite{matsu10}, such cumulants can be found
as weighted marginals of the underlying bispectrum, (to third order), 
trispectrum (to fourth order), {\sl etc.}. 
On a sphere, the high order marginals are particularly cumbersome and time
consuming to compute, as they also involve the contractions of $n-j$ Wigner
symbols. Here we suggest a different route,
based on the assumption that scientists interested in fitting extrema counts to
non-Gaussian maps are
 typically in a position to generate realizations of such maps.  In that case,
it becomes relatively straightforward to draw samples of such maps, and
estimate the corresponding cumulants. 
  The {\sc \small HEALPix} \citep{hivon05} library provides in fact a direct estimate of the derivatives of such maps up to second order, which is all that is required to compute  the  cumulants of the JPDF.

As an illustration, let us generate sets of parameterized non-Gaussian maps
using the package {\tt sky-ng-sim} \cite{Rocha} of  {\sc \small HEALPix}. 
In this so called harmonic model, the PDF of the pixel temperature, $T$ is given by
 $\exp(-T^2/2 \sigma_0^2) \left| \sum_{i=0}^n \alpha_i C_i 
H_i(T/\sigma_0)\right|^2$, 
where $C_i$ are normalization constants.  In this paper, we use 
 {\sc \small nside}=2048, $\ell_{\max}=4096$, $n=2$, $\sigma_0=1$, $\alpha_{0}=0$ and
 vary $\alpha_1$ and  $\alpha_2$.    We also consider the second option of {\tt sky-ng-sim} which produces 
 non Gaussian field as even power, $\beta$ of  unit variance zero mean Gaussian fields.
For each set of maps, we compute its derivatives, and arithmetically  average
 the corresponding cumulants, using a code, {\tt map2cum}
relying on the  {\sc \small HEALPix} routine {\tt alm2map\_der}.
Invariant variables $J_1$ and $J_2$  on a sphere are
defined via the mixed tensor of covariant derivatives $J_1 = {x_{;i}}^{;i}$ and
$J_2 = \left| {x_{;i}}^{;j} \right|$.
The differential counts  are then evaluated for a range of 
threshold, $\nu\in[-5,5]$. For each of these maps, the number of extrema is
computed by the procedure {\tt map2ext}  which implements the following algorithm:
for every pixel a segment of quadratic surface is 
fit in the tangent plane  based on the
temperature values at the pixel of origin and its  {\sc \small HEALPix} 
neighbours. The position of the extremum of this quadratic, its height and
its Hessian are computed. The extremum is counted into the tally of the type
determined by its Hessian if its position falls within the original pixel.
Several additional checks are performed to preclude registering extrema in the
neighbouring pixels and minimize missing extrema
due to jumps in the fit parameters as region shifts to the next pixel.
Masks are treated by not considering pixels next to the mask boundary.
Pixel-pixel noise covariance can be included while doing the local fit.
On noise-free maps the procedure performs with better than 1\% accuracy when the
map is smoothed with Gaussian filter with FWHM exceeding 6 pixels. Both 
{\tt map2cum} and {\tt map2ext} are available upon request.
Figure~\ref{fig:mainfig} illustrates the very good agreement between the theoretical
 expectation of the differential number counts to the measured ones for both the harmonic and the power-law models.

An alternative numerical procedure, which is likely to be more practical for expansion beyond the fourth order  was also successfully explored for 2D topological invariants.
Starting from Eq.~(\ref{eq:2DP_general}), we re-express both the polynomials in $J_1,J_2,\zeta$, and $q^2$ and $G_{\rm 2D}$ in terms of the six field variables,
$(x,x_i,x_{ij})$. We then construct formally the marginal $G_{\nu}(\mathbf{x}=(x_{11},x_{12},x_{22})| x=\nu, x_1=x_2=0)$, where the latter condition corresponds to imposing that 
we are seeking extrema of the field.
It becomes straightforward  
to draw large sets of 3 random numbers satisfying $G_{\nu}$.  For each triplets, $\mathbf{x}$, 
 and a given numerical set of cumulants, 
we then compute the argument, ${\cal I}(\mathbf{x})$ of the square bracket in  Eq.~(\ref{eq:2DP_general}) (up to some given order),   together with the two eigenvalues of the Hessian. 
For maxima (resp. minima, resp. saddle points), we replace  ${\cal I}$ by 0 if the two eigenvalues are not negative (resp. positive, resp. of different sign). 
The sum over all triplets yields a Monte Carlo estimate of $\partial n_{\rm ext}/\partial \nu$. The accuracy of the estimate depends on the extent of rejection while 
applying the extremal condition. 

Note in closing that all the presented analysis is straightforwardly generalized to 3D (noticeably the Monte Carlo method), as shown in \citep{GPP},
to describe the large scale distribution of matter.  Indeed in this context,
the gravitational instability that nonlinearly maps the initial Gaussian 
inhomogeneities in matter density into the LSS, 
induces strong non-Gaussian features culminating in the formation of
collapsed, self-gravitating objects such as galaxies and clusters of galaxies.
\\
\\
{\sl Acknowledgments:}
we warmly thank E. Hivon for his help. CP and DP thanks the department of physics, Oxford, for hospitality during the completion of this work. The Gram-Charlier to cumulants lookup table is available at  {\tt http://www.iap.fr/users/pichon/Gram/},
together with the second order extrema counts, and third order genus given by
$K_1$. 
All codes to compute the cumulants of given fields and the extrema on the  {\sc \small HEALPix}  pixelisation
are available upon request.

\bibliography{dummy}

\begin{thebibliography}{10}
\expandafter\ifx\csname natexlab\endcsname\relax\def\natexlab#1{#1}\fi
\expandafter\ifx\csname bibnamefont\endcsname\relax
  \def\bibnamefont#1{#1}\fi
\expandafter\ifx\csname bibfnamefont\endcsname\relax
  \def\bibfnamefont#1{#1}\fi
\expandafter\ifx\csname citenamefont\endcsname\relax
  \def\citenamefont#1{#1}\fi
\expandafter\ifx\csname url\endcsname\relax
  \def\url#1{\texttt{#1}}\fi
\expandafter\ifx\csname urlprefix\endcsname\relax\def\urlprefix{URL }\fi
\providecommand{\bibinfo}[2]{#2}
\providecommand{\eprint}[2][]{\url{#2}}

\bibitem[{\citenamefont{{Adler}}(1981)}]{Adler}
\bibinfo{author}{\bibfnamefont{R.~J.} \bibnamefont{{Adler}}},
  \emph{\bibinfo{title}{The Geometry of Random Fields}}
  (\bibinfo{publisher}{The Geometry of Random Fields, Chichester: Wiley},
  \bibinfo{year}{1981}).

\bibitem[{\citenamefont{{Doroshkevich}}(1970)}]{Doroshkevich}
\bibinfo{author}{\bibfnamefont{A.~G.} \bibnamefont{{Doroshkevich}}},
  \bibinfo{journal}{Astrofizika} \textbf{\bibinfo{volume}{6}},
  \bibinfo{pages}{581} (\bibinfo{year}{1970}).

\bibitem[{\citenamefont{{Bardeen} et~al.}(1986)\citenamefont{{Bardeen}, {Bond},
  {Kaiser}, and {Szalay}}}]{BBKS}
\bibinfo{author}{\bibfnamefont{J.~M.} \bibnamefont{{Bardeen}}},
  \bibinfo{author}{\bibfnamefont{J.~R.} \bibnamefont{{Bond}}},
  \bibinfo{author}{\bibfnamefont{N.}~\bibnamefont{{Kaiser}}}, \bibnamefont{and}
  \bibinfo{author}{\bibfnamefont{A.~S.} \bibnamefont{{Szalay}}},
  \bibinfo{journal}{\apj} \textbf{\bibinfo{volume}{304}}, \bibinfo{pages}{15}
  (\bibinfo{year}{1986}).

\bibitem[{\citenamefont{{Pogosyan} et~al.}(2009)\citenamefont{{Pogosyan},
  {Gay}, and {Pichon}}}]{PGP}
\bibinfo{author}{\bibfnamefont{D.}~\bibnamefont{{Pogosyan}}},
  \bibinfo{author}{\bibfnamefont{C.}~\bibnamefont{{Gay}}}, \bibnamefont{and}
  \bibinfo{author}{\bibfnamefont{C.}~\bibnamefont{{Pichon}}},
  \bibinfo{journal}{\prd} \textbf{\bibinfo{volume}{80}},
  \bibinfo{pages}{081301} (\bibinfo{year}{2009}), \eprint{0907.1437}.

\bibitem[{\citenamefont{{Matsubara}}(2003)}]{Matsubara}
\bibinfo{author}{\bibfnamefont{T.}~\bibnamefont{{Matsubara}}},
  \bibinfo{journal}{\apj} \textbf{\bibinfo{volume}{584}}, \bibinfo{pages}{1}
  (\bibinfo{year}{2003}).

\bibitem[{\citenamefont{{Matsubara}}(2010)}]{matsu10}
\bibinfo{author}{\bibfnamefont{T.}~\bibnamefont{{Matsubara}}},
  \bibinfo{journal}{\prd} \textbf{\bibinfo{volume}{81}},
  \bibinfo{pages}{083505} (\bibinfo{year}{2010}), \eprint{1001.2321}.

\bibitem[{\citenamefont{{Longuet-Higgins}}(1957)}]{Longuet}
\bibinfo{author}{\bibfnamefont{M.~S.} \bibnamefont{{Longuet-Higgins}}},
  \bibinfo{journal}{Royal Society of London Philosophical Transactions Series
  A} \textbf{\bibinfo{volume}{249}}, \bibinfo{pages}{321}
  (\bibinfo{year}{1957}).

\bibitem[{\citenamefont{{G{\'o}rski} et~al.}(2005)\citenamefont{{G{\'o}rski},
  {Hivon}, {Banday}, {Wandelt}, {Hansen}, {Reinecke}, and
  {Bartelmann}}}]{hivon05}
\bibinfo{author}{\bibfnamefont{K.~M.} \bibnamefont{{G{\'o}rski}}},
  \bibinfo{author}{\bibfnamefont{E.}~\bibnamefont{{Hivon}}},
  \bibinfo{author}{\bibfnamefont{A.~J.} \bibnamefont{{Banday}}},
  \bibinfo{author}{\bibfnamefont{B.~D.} \bibnamefont{{Wandelt}}},
  \bibinfo{author}{\bibfnamefont{F.~K.} \bibnamefont{{Hansen}}},
  \bibinfo{author}{\bibfnamefont{M.}~\bibnamefont{{Reinecke}}},
  \bibnamefont{and}
  \bibinfo{author}{\bibfnamefont{M.}~\bibnamefont{{Bartelmann}}},
  \bibinfo{journal}{\apj} \textbf{\bibinfo{volume}{622}}, \bibinfo{pages}{759}
  (\bibinfo{year}{2005}), \eprint{arXiv:astro-ph/0409513}.

\bibitem[{\citenamefont{{Rocha} et~al.}(2005)\citenamefont{{Rocha}, {Hobson},
  {Smith}, {Ferreira}, and {Challinor}}}]{Rocha}
\bibinfo{author}{\bibfnamefont{G.}~\bibnamefont{{Rocha}}},
  \bibinfo{author}{\bibfnamefont{M.~P.} \bibnamefont{{Hobson}}},
  \bibinfo{author}{\bibfnamefont{S.}~\bibnamefont{{Smith}}},
  \bibinfo{author}{\bibfnamefont{P.}~\bibnamefont{{Ferreira}}},
  \bibnamefont{and}
  \bibinfo{author}{\bibfnamefont{A.}~\bibnamefont{{Challinor}}},
  \bibinfo{journal}{\mnras} \textbf{\bibinfo{volume}{357}}, \bibinfo{pages}{1}
  (\bibinfo{year}{2005}), \eprint{arXiv:astro-ph/0406136}.

\bibitem[{\citenamefont{{Gay} et~al.}(2011)\citenamefont{{Gay}, {Pichon}, and
  {Pogosyan}}}]{GPP}
\bibinfo{author}{\bibfnamefont{C.}~\bibnamefont{{Gay}}},
  \bibinfo{author}{\bibfnamefont{C.}~\bibnamefont{{Pichon}}}, \bibnamefont{and}
  \bibinfo{author}{\bibfnamefont{D.}~\bibnamefont{{Pogosyan}}},
  \bibinfo{journal}{in prep.}  (\bibinfo{year}{2011}).

\end{thebibliography}
\end{document}